\documentclass[twocolumn,comment]{jpsj3}
\usepackage{graphicx}

\title{Comment on "Low-Temperature Magnetic Properties of the Kondo Lattice Model in One Dimension"}

\author{Hiroaki Matsueda\thanks{matsueda@sendai-nct.ac.jp}}
\inst{Sendai National College of Technology, Sendai 989-3128, Japan} %\\

\begin{document}

\maketitle

In Ref.~\citen{Minami}, incommensurate magnetic structures of the one-dimensional (1D) Kondo-lattice model were re-examined. This model has been already examined in a context of perovskite manganise oxides~\cite{Yunoki1,Yunoki2}, but the main issue here is about the difference between the Kondo-lattice and RKKY models in a small Kondo-coupling region. The authors found collinear spin configurations as well as spiral and noncoplanar phases. In particular, the four-period collinear phase appears for a very small coupling. However this is not seen in the RKKY model, although the RKKY model can be derived from the perturbation for the coupling. The authors attributed this special situation to the energy decrease by the gap opening in the electron band dispersion. In this comment, we perform a microscopic calculation of the band structure to understand the gap opening in a case of four-period collinear spin configuration.

We start with the 1D Kondo-lattice model. The Hamiltonian is given by
\begin{eqnarray}
H = -\sum_{i,j}t_{ij}c_{i}^{\dagger}c_{j} - \frac{J}{2}\sum_{i}c_{i}^{\dagger}\vec{\sigma}c_{i}\cdot\vec{S}_{i},
\end{eqnarray}
where $c_{i}$ is spinor representation of an electron annihilation operator at site $i$, $\vec{S}$ is a classical vector spin with a normalization condition $\vec{S}_{i}\cdot\vec{S}_{i}=1$, $t_{ij}=t\alpha_{ij}$, and $\alpha_{ij}=\delta_{i,j+1}+\delta_{i,j-1}$. Here, the ferromagnetic Kondo coupling $J$ is introduced, and for large $J$ values this term induces ferromagnetism by the double-exchange mechanism. Here, there is no quantum dynamics for the classical spin, i.e. $[\vec{S}_{i},H]=0$.

We examine the single-particle propagator for electron. The propagator can be formally represented as
\begin{eqnarray}
G(\omega,k)=\frac{1}{\omega-E(k)-\Sigma(\omega,k)},
\end{eqnarray}
where $E(k)$ is the dispersion relation for non-interacting electrons and $\Sigma(\omega,k)$ is the self-energy correction. The equation of motion for the electron operator in the Heisenberg representation is given by
\begin{eqnarray}
i\frac{\partial}{\partial\tau}c_{i}(\tau)=-tc_{i}^{\alpha}(\tau)-\frac{J}{2}\vec{S}_{i}\cdot\vec{\sigma}c_{i}(\tau),
\end{eqnarray}
where $c^{\alpha}_{i}=\sum_{j}\alpha_{ij}c_{j}$. This equation of motion gives explicit forms of $E(k)$ and $\Sigma(\omega,k)$. The non-interacting dispersion $E(k)$ is given by $E_{\pm}(k)=-t\alpha(k)\mp(J/2)m$, where the sign $\pm$ denote the up or down spin component and $m$ is the expectation value of the local magnetic moment for the up spin. Since we will focus on the collinear spin phase, we can take $m=0$ and $E(k)=-2t\cos k$. The self-energy correction is given by
\begin{eqnarray}
\Sigma(x-y)=\left(\frac{J}{2}\right)^{2}\left<R\vec{S}_{i}\cdot\vec{\sigma}c_{i}(\tau)c_{j}^{\dagger}(\tau^{\prime})\vec{\sigma}\cdot\vec{S}_{j}\right>_{I},
\end{eqnarray}
where $x=(i,\tau)$ and $y=(j,\tau^{\prime})$, $R$ and $I$ denote the retarded and irreducible parts, respectively. The Fourier transform of $\Sigma$ can be represented as
\begin{eqnarray}
\Sigma(\omega,k)=\left(\frac{J}{2}\right)^{2}\sum_{q}\chi(q)G(\omega,k-q),
\end{eqnarray}
where $\chi(q)$ is static spin susceptibility. We can thus obtain the propagator by iteration numerically, if the susceptibility is given in advance. When a particular spin correlation characterized by $q=q_{0}$ is enhanced, the correction is simply given by $\Sigma(\omega,k)=(J/2)^{2}\left(G(\omega,k-q_{0})+G(\omega,q+q_{0})\right)/2$ with $\chi(q_{0})=\chi(-q_{0})=1/2$.

The self-energy correction for double-exchange ferromagnetic case ($q=0$) is given by $\Sigma(\omega,k)=(J/2)^{2}G(\omega,k)$. In this case, the damping of the non-interacting spectrum at $\omega=E(k)$ is characterized by $\Sigma\left(\omega=E(k),k\right)=(J/2)i$, and this just gives a uniform broadening to the non-interacting dispersion. Then, the mixing gap does not open, and the simple metallic band with a finite life time remains. On the other hand, the correction for four-period states ($q=\pi/2$) is given by $\Sigma(\omega,k)=(J/2)^{2}\left(G(\omega,k-\pi/2)+G(\omega,k+\pi/2)\right)/2$, and then the momentum transfer mixes the different states, leading to the gap opening.

\begin{figure}[htbp]
\begin{center}
\includegraphics[width=6.5cm]{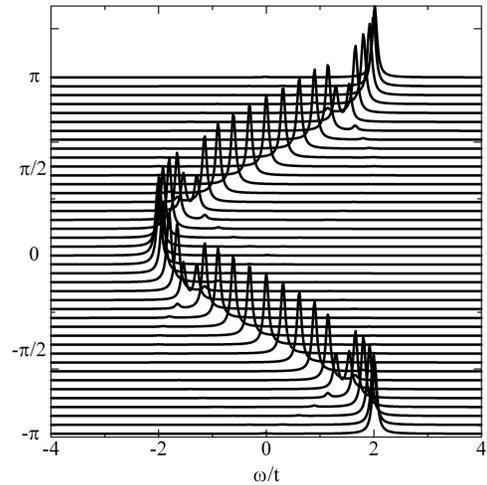}
\end{center}
\caption{Angle-resolved spectrum $A(\omega,k)=(-1/\pi)\Im G(\omega+i\epsilon,k)$ for $J=0.25t$ and $\epsilon=0.05t$.}
\label{fig1}
\end{figure}

We perform a self-consistent calculation to obtain the imaginary part of this propagator. Figure~\ref{fig1} shows the angle-resolved spectrum for $J=0.25t$. We actually observe the gap at around $k=\pm\pi/4$, and the system is insulating for $1/4$-filling. The gap size is roughly $J$, and thus the perturbation approach breaks out even for infinitesimal $J$ values. We have also observed the gap at $k=\pm 3\pi/4$, which is also consistent with Ref.~\citen{Minami}.

It is important to see how electrons are dressed with spin fluctuation to form the gapped band structure. For this purpose, let us consider the equations of motion for higher-order composite excitations. The second-order equation of motion is given by 
\begin{eqnarray}
i\frac{\partial}{\partial\tau}\vec{S}_{i}\cdot\vec{\sigma}c_{i}(\tau) = -t\vec{S}_{i}\cdot\vec{\sigma}c_{i}^{\alpha}(\tau) - \frac{J}{2}c_{i}(\tau),
\end{eqnarray}
and the third-order equation is also given by
\begin{eqnarray}
i\frac{\partial}{\partial\tau}\vec{S}_{i}\cdot\vec{\sigma}c_{i}^{\alpha}(\tau) &=& -2t\vec{S}_{i}\cdot\vec{\sigma}c_{i}(\tau) - t\vec{S}_{i}\cdot\vec{\sigma}c_{i}^{\alpha^{\prime}}(\tau) \nonumber \\
&& -\frac{J}{2}\vec{S}_{i}\cdot\vec{\sigma}\bigl(\vec{S}_{i}\cdot\vec{\sigma}c_{i}(\tau)\bigr)^{\alpha},
\end{eqnarray}
where $\alpha_{ij}^{\prime}=\delta_{i,j-2}+\delta_{i,j+2}$. The equation of motion for $\vec{S}\cdot\vec{\sigma}c^{\alpha^{\prime}}$ is given by
\begin{eqnarray}
i\frac{\partial}{\partial\tau}\vec{S}_{i}\cdot\vec{\sigma}c_{i}^{\alpha^{\prime}}(\tau) &=& -t\vec{S}_{i}\cdot\vec{\sigma}c_{i}^{\alpha}(\tau) - t\vec{S}_{i}\cdot\vec{\sigma}c_{i}^{\alpha^{\prime\prime}}(\tau) \nonumber \\
&& - \frac{J}{2}\vec{S}_{i}\cdot\vec{\sigma}\bigl(\vec{S}_{i}\cdot\vec{\sigma}c_{i}(\tau)\bigr)^{\alpha^{\prime}},
\end{eqnarray}
where $\alpha_{ij}^{\prime\prime}=\delta_{i,j-3}+\delta_{i,j+3}$.

According to the equations of motion, the relevant basis states are given by
\begin{eqnarray}
\psi=\left(
\begin{array}{c}
\psi^{1} \\
\psi^{2} \\
\psi^{3} \\
\psi^{4} \\
\psi^{5} \\
\psi^{6} \\
\psi^{7} \\
\vdots
\end{array}
\right)
=\left(
\begin{array}{c}
c \\
\vec{S}\cdot\vec{\sigma}c \\
\vec{S}\cdot\vec{\sigma}c^{\alpha} \\
\vec{S}\cdot\vec{\sigma}c^{\alpha^{\prime}} \\
\vec{S}\cdot\vec{\sigma}\bigl(\vec{S}\cdot\vec{\sigma}c\bigr)^{\alpha} \\
\vec{S}\cdot\vec{\sigma}c^{\alpha^{\prime\prime}} \\
\vec{S}\cdot\vec{\sigma}\bigl(\vec{S}\cdot\vec{\sigma}c\bigr)^{\alpha^{\prime}} \\
\vdots
\end{array}
\right).
\end{eqnarray}
The single-particle propagator $G(\omega,k)$ is defined as the Fourier transform of the $(1,1)$-component of the retarded propagator matrix $\mathcal{G}(x-y)=\bigl<R\psi(x)\psi^{\dagger}(y)\bigr>$. The Fourier transform of $\mathcal{G}(x-y)$ is given by
\begin{eqnarray}
\mathcal{G}(\omega,k)=I(k)\frac{1}{\omega I-M(k)-\delta M(\omega,k)}I(k).
\end{eqnarray}
The normalization matrix $I$, energy matrix $M$, and dynamical correction $\delta M$ are respectively given by
\begin{eqnarray}
&& I(x-y)=\delta(\tau-\tau^{\prime})\left<\left\{\psi(x),\psi^{\dagger}(y)\right\}\right>, \\
&& M(x-y)=\delta(\tau-\tau^{\prime})\left<\left\{i\frac{\partial}{\partial\tau}\psi(x),\psi^{\dagger}(y)\right\}\right>, \\
&& \delta M(x-y)=\left<R i\frac{\partial}{\partial\tau}\psi(x) \left(i\frac{\partial}{\partial\tau}\psi(y)\right)^{\dagger}\right>_{I}.
\end{eqnarray}
They are Hermitian matrices, $I=I^{\dagger}$, $M=M^{\dagger}$, and $\delta M=\delta M^{\dagger}$, respectively.

We evaluate the first $4\times 4$-compoments of $I$ and $M$. For the four-period collinear spin case ($\uparrow\uparrow\downarrow\downarrow\uparrow\uparrow\downarrow\downarrow\cdots, q=\pi/2$), we can take $\bigl<\vec{S}_{i}\cdot\vec{S}_{i+n}\bigr>=\Re e^{i\pi n/2}$ ($n=0,1,2,3$). We find
\begin{eqnarray}
&& I^{11}(k) = I^{22}(k) = 1 \; , \; I^{33}(k) = 2\left(1-\cos(2k)\right) , \\
&& I^{24}(k) = I^{42}(k) = -2\cos(2k) , \\
&& I^{44}(k)= 2\left(1+\cos(4k)\right) ,
\end{eqnarray}
(the other components are all zero) and
\begin{eqnarray}
&& M^{1a}(k) = -t\alpha(k)I^{1a}(k) - \frac{J}{2}I^{2a}(k) , \\
&& M^{2a}(k) = -tI^{3a}(k) - \frac{J}{2}I^{1a}(k) , \\
&& M^{3a}(k) = -2tI^{2a}(k) -tI^{4a}(k) - \frac{J}{2}I^{5a}(k) , \\
&& M^{44}(k) = -tI^{34}(k) - tI^{64}(k) - \frac{J}{2}I^{74}(k) ,
\end{eqnarray}
where $a=1,2,3,4$ and $I^{53}=I^{54}=I^{64}=I^{74}=0$. Note that higher-order anti-commutation relations induce several chirality operators such as $\gamma_{i}=\vec{S}_{i-1}\cdot\vec{S}_{i}\times\vec{S}_{i+1}=\epsilon_{\mu\nu\lambda}S_{i-1}^{\mu}S_{i}^{\nu}S_{i+1}^{\lambda}$, and they should have locally finite amplitudes in the incommensurate phase. However, their averages are zero, and basically they do not change the dispersion.

The unrenormalized bands before their mixing due to the off-diagonal matrix elements can be defined by $\epsilon_{\mu}(k)=M^{\mu\mu}(k)/I^{\mu\mu}(k)$ and
\begin{eqnarray}
\epsilon_{1}(k)=-t\alpha(k) \; , \; \epsilon_{2}(k)=\epsilon_{3}(k)=\epsilon_{4}(k)=0 .
\end{eqnarray}
We find that the unrenormalized bands are all flat except for $\epsilon_{1}(k)$. Since the off-diagonal coupling is momentum-dependent, the mixing rule is somehow special. At $k=\pi/4$, we find that the primary band couples with only two composite bands, since $I^{44}$ becomes zero as
\begin{eqnarray}
I(\pi/4)=\left(\begin{array}{cccc}1&0&0&0\\0&1&0&0\\0&0&2&0\\0&0&0&0\end{array}\right).
\end{eqnarray}
This feature protects the broadening of the gap by much higher-order contributions. In this case, the propagator matrix is given by
\begin{eqnarray}
\mathcal{G}(\omega,\pi/4)=\left(\begin{array}{ccc}\omega+\sqrt{2}t&J/2&0\\J/2&\omega&t\\0&t&\omega/2\end{array}\right)^{-1}, \label{mathcalG}
\end{eqnarray}
and the two main poles across the gap for $J=0.25t$ are located at $\omega=-1.51t$ and $\omega=-1.33t$. Their energy difference is $0.18t$ roughly comparable with the $J$ value. If we only use two poles, the gap size is highly overestimated. This is because that the off-diagonal coupling between $\psi^{2}$ and $\psi^{3}$ in Eq.~(\ref{mathcalG}) is of order $t$.

Summarizing, we have examined the detailed band structure of the 1D Kondo lattice model. For the four-period collinear spin configuration, the band gap opens owing to the mixing among primary and higher-order composite bands. This feature brings some trouble to a simple perturbation approach.

\end{document}